\documentclass[prd,aps,showpacs,showkeys,twocolumn,10pt,nofootinbib]{revtex4-1}
\usepackage{amssymb,amsmath,amsthm}
\usepackage{mathrsfs}
\usepackage[applemac]{inputenc}
\usepackage{color}

\begin{document}

\title{$f(R)$ quantum cosmology: avoiding the Big Rip}

 \author{Ana Alonso-Serrano} 
\email{ana.alonso.serrano@aei.mpg.de}
\affiliation{Max Planck Institute for Gravitational Physics, Albert Einstein Institute, Am M\"uhlenberg 1, D-14476 Postdam-Golm, Germany}
\affiliation{Institute of Theoretical Physics, Faculty of Mathematics and Physics, Charles University, 18000 Prague, Czech Republic}

\author{Mariam Bouhmadi-L\'opez}
\email{mariam.bouhmadi@ehu.eus}
\affiliation{Department of Theoretical Physics, University of the Basque Country
UPV/EHU, \\P.O. Box 644, 48080 Bilbao, Spain}
\affiliation{IKERBASQUE, Basque Foundation for Science, 48011, Bilbao, Spain}
\author{Prado Mart\'in-Moruno}
\email{pradomm@ucm.es}
\affiliation{Departamento de F\'isica Te\'orica and UPARCOS, Universidad Complutense de Madrid, E-28040 Madrid, Spain}

\date{\today}

\begin{abstract}
Extended theories of gravity have gathered a lot of attention over the last years, for they not only provide an excellent framework to describe the inflationary era but also yields an alternative to the elusive and mysterious dark energy. Among the different extended theories of gravity, on this work we focus on metric $f(R)$ theories. In addition, it is well known that if the late-time acceleration of the universe is stronger than the one induced by a cosmological constant then some future cosmic singularities might arise, being the Big Rip the most virulent one. Following this reasoning, on this work, we analyse the Big Rip singularity in the framework of $f(R)$ quantum geometrodynamics. Invoking the DeWitt criterion, i.~e.~that the wave function vanishes at the classical singularity, we proof that a class of solutions to the Wheeler--DeWitt equation fulfilling this condition can be found. Therefore, this result hints towards the avoidance of the Big Rip in metric $f(R)$ theories of gravity.
\end{abstract}


\maketitle

\section{Introduction}

Nowadays there is no doubt that our Universe is currently undergoing a phase of accelerated expansion \cite{Ade:2015xua}. This acceleration can be described in the framework of General Relativity (GR) assuming the existence
of dark energy, which is a fluid violating the strong energy condition and leading only to gravitational effects. Dark energy could well be originated by a cosmological constant, usually interpreted as a vacuum
energy, if we accept that we do not know how to calculate its value \cite{Sahni:1999gb,Carroll:2000fy,Weinberg:1988cp,Padmanabhan:2002ji}.
On the other hand, soon after the discovery of the accelerated expansion of our Universe, it was emphasised that dark energy could lead even to more acceleration than a universe whose dynamics is driven by a 
cosmological constant \cite{Caldwell:1999ew}.
This is the case for a phantom fluid, a class of dark energy that violates even the null energy condition and, therefore, has an energy density which grows with the cosmic expansion.
Moreover, recent observational data continue to be fully compatible with the possibility that phantom energy is driving the dynamics of our Universe \cite{Ade:2015xua}

It is already well known that phantom energy may lead to the occurrence of a Big Rip (BR) singularity \cite{Caldwell:1999ew,Starobinsky:1999yw,Caldwell:2003vq}. 
This doomsday corresponds to a curvature singularity characterised by a divergence at a finite cosmic time and infinite scale factor of both the Hubble parameter and its cosmic time derivative, implying the divergence of the phantom energy density and pressure in a general relativistic  framework. Nevertheless, this is not necessarily always the case as phantom cosmologies could have an asymptotic de Sitter behaviour~\cite{BouhmadiLopez:2004me}. Moreover, phantom energy could also lead to the occurrence of a Big Freeze singularity
\cite{BouhmadiLopez:2006fu,BouhmadiLopez:2007qb}, which is characterised by a divergence of the Hubble parameter and its cosmic time derivative but at a finite and non-vanishing scale factor. These investigations have renewed the interest in studying new cosmic singularities, which are not necessarily due to the existence of a phantom fluid, and there is a whole bunch of them (for a recent account on this topic, please see reference \cite{MCP}).

When the Universe evolves towards (from) a future (past) singularity, the gravitational theory cannot provide us further physical information. In fact what happens is that the structure of  classical spacetime is broken at those curvature singularities
and, therefore, a metric theory of gravity is not well defined any more. Hence we need to resort to a quantum gravitational theory in order to unveil the final fate (if any) of the universe. There are different candidates for such a quantum gravitational theory that come from very different approaches. We are interested in what is known as canonical quantum gravity, that is a non-perturbative and background independent quantisation. In this context, we will use the metric variables as configuration space in what was called quantum geometrodynamics, that also provides the well-known expression of Wheeler--DeWitt equation \cite{Wheeler,DeWitt}. Although this theory is not solvable in a general context (as none of the other proposals), it provides some interesting tools and results that can be implemented in simpler (minisuperspace) cosmological models \cite{Halliweell,Misner,Kuchar}, preserving some characteristics of the full theory, which can provide us with helpful information in order to explore some quantum concepts in cosmology \cite{clausbook}. Within this approach, one of the common and interesting studies is the one related to the analysis of singularities, as we expect that a quantum theory of gravity to be able to solve or at least appease the issue of singularities.

On the other hand, the current accelerated expansion of our Universe could be a signal of the unsuitability of GR at cosmological scales already at the classical level. Indeed, it has been argued that GR has not being  properly tested at the strong-field regime and its validity at cosmological scales is simply assumed. Therefore, alternative theories of gravity have acquired a renewed interest as potential candidates to describe the physical phenomena in our Universe. Some of those theories, as extended theories of gravity \cite{Capozziello:2011et}, had been motivated in the past as effective theories of gravity that may encapsulate some semi-classical effects coming from
the underlying quantum gravitational framework. Metric $f(R)$ theories of gravity, which assumes a gravitational Lagrangian which depends on a function of the scalar curvature, are probably the most studied theories of this kind \cite{Sotiriou:2008rp,Nojiri:2010wj}.
It should be emphasised, however, that the current phenomenological motivation for considering these theories as viable gravitational theories is not necessarily always based on that effective approach. Following this spirit, one could understand extended theories of gravity as classical fundamental theories able to describe the current cosmological phase without the introduction of dark fluids.

As alternative theories of gravity can describe the same background cosmological evolution as GR, the same kind of cosmological singularities crop up also in these scenarios. Those singularities signal the need of considering a quantum formulation of alternative theories of gravity (assumed as fundamental) as has been recently carried out, for example, in Palatini Eddington-Born-Infeld theories \cite{Bouhmadi-Lopez:2016dcf,Albarran:2017swy}.
Given that metric $f(R)$ gravity is one of the simplest alternative theories of gravity, it is especially interesting to consider these theories as proxy theories to investigate the potential quantum avoidance of the BR singularity when such a singularity is completely due to the modified Hilbert-Einstein action. Therefore, in this paper we will consider the formulation of $f(R)$ quantum cosmology to analyse the quantum fate  of the universe close to a BR singularity using a geometrodynamic formulation of quantum cosmology. In fact, we will use the metric variables to construct the configuration space within the above geometrodynamics approach that has the advantage of recovering the correct semi-classical limit \cite{clausbook}. In this scenario, we will analyse the behaviour of the BR in an $f(R)$ metric theory using the so-called DeWitt criterion (DW) that establish that the singularity is potentially avoided if the wavefunction vanishes in the configuration space in that region~\cite{DeWitt}.

This paper can be outlined as follows: In section \ref{sec:br}, we summarise some basic results regarding the BR singularity and its occurrence in $f(R)$-gravity. In section \ref{sec:QC}, we recapitulate
some old results regarding a suitable formulation of $f(R)$ quantum cosmology, adapting them to our needs. In section \ref{sec:QfR}, we particularise the modified Wheeler--DeWitt (WDW) equation of the 
$f(R)$ quantum cosmology to a theory leading to a BR singularity to investigate the behaviour of the wave function of the universe close to this singularity. We summarise and present our conclusion in section \ref{sec:conc.}.  Finally, in appendix \ref{A1}, we proof the suitability of the approximations we used when getting the wave functions that fulfils the the modified WDW equation.

\section{The Big Rip in $f(R)$-cosmology}\label{sec:br}

As it is well known, phantom fluids can lead to the occurrence of future singularities in a general relativistic background.
If phantom energy is characterised by a constant equation of state parameter, $w=p/\rho<-1$, a BR singularity will take place,
which is a curvature singularity at which both the scale factor and the energy density diverge.
One can easily find this singularity noting that the scale factor evolution of a universe filled only with that phantom fluid is
\begin{equation}\label{a_phantom}
 a(t)=a_\star (t_{\textrm{br}}-t)^{-\frac{2}{3(|w|-1)}},
\end{equation}
where $a_\star$ is an integration constant, which can be expressed as $a_\star=a_0[3C(|w|-1)/2]^{-2/[3(|w|-1)]}$, $C=(\kappa\,\rho_0/3)^{1/2}$, $\kappa=8\pi G$,  the sub-index $0$ denotes evaluation at $t_0$, 
and $t_{\textrm{br}}$ correspond to the cosmic time when the big rip takes place.
This is
\begin{equation}\label{tbr}
 t_{\rm br}=t_0+\frac{2}{3\,C(|w|-1)}>t_0.
\end{equation}
It should be emphasised that this phantom model could describe the late-time evolution of our Universe when the matter content will be diluted by the cosmological expansion.
Moreover, this model is not the only one that can lead to a BR future singularity, but it is probably the simplest one.

On the other hand, metric $f(R)$-theories of gravity are described by a gravitational action of the form
\begin{equation}\label{L1-1}
 S=\int d^4x\sqrt{-g}f(R),
\end{equation}
where each $f(R)$ corresponds to a different theory, being $f(R)=R/(2\kappa)$ just GR. In cosmological scenarios, one can write this action as
\begin{equation}
 S=\int dt\,\mathcal{L}(a,\,\dot a,\,\ddot a),
\end{equation}
with
\begin{equation}\label{L1}
 \mathcal{L}(a,\,\dot a,\,\ddot a)=\mathcal{V}_{(3)}\,a^3f(R),
\end{equation}
where $\mathcal{V}_{(3)}$ is the spatial 3-volume and, for simplicity, we have chosen a lapse function $N=1$. 
Assuming a matter action of a perfect fluid with energy density $\rho$ and pressure $p$ added to the gravitational action (\ref{L1}), one can obtain the following modified Friedmann and Raychaudhuri equations:
\begin{eqnarray}
	\label{eq: Fried1}
	H^2 &=& \frac{1}{6f_R} \left( \rho - f+f_RR-6H f_{RR}\dot{R}\right),\\
\label{eq: Fried2}
	2\dot{H} + 3H^2 &=& \frac{1}{2f_R} \left[ p + f-f_RR \right. \nonumber \\
	&+&\left. 2\left(f_{RRR}\dot{R}^2 +f_{RR}\ddot{R}+2Hf_{RR}\dot{R}\right)\right],
\end{eqnarray}
where $f_R\equiv df/dR$, $f_{RR}\equiv d^2f/dR^2$, and $f_{RRR}\equiv d^3f/dR^3$, and the dependence of $f$ and its derivatives on $R$ is not explicitly stated.

Any general relativistic background cosmology can be reconstructed in the context of $f(R)$-theories by choosing an appropriate $f(R)$ function. 
Therefore, $f(R)$-gravity may lead to all four types of cosmic singularities that can appear in GR \cite{Nojiri:2008fk,Bamba:2008ut,Capozziello:2009hc,Nojiri:2017ncd}.
In particular, in reference 
\cite{Morais:2015ooa} the authors showed that the cosmological evolution generated by a fluid with a constant equation of state parameter $w$ in a general relativistic cosmological scenario
can be described in an $f(R)$ theory with \cite{Morais:2015ooa}
\begin{equation}
f(R)=C_+R^{\beta_+}+C_-R^{\beta_-},
\end{equation}
where
\begin{eqnarray}\label{beta}
\beta_{\pm}&=&\frac{1}{2}\left[1+\frac{1+3w}{6(1+w)}\right. \nonumber\\
&\pm& \left. \sqrt{\frac{2(1-3w)}{3(1+w)}+\left[1+\frac{1+3w}{6(1+w)}\right]^2}\right],
\end{eqnarray}
and $C_+$ and $C_-$ are arbitrary constants.
If we want to describe a phantom model with the evolution given by equation (\ref{a_phantom}), it is important to note that for values $w <-1$, $\beta_{\pm}$ is complex valued, 
that is, it has the form $\beta_{\pm}=\gamma+i\sigma$. Hence, we can write:
\begin{equation}
f(R)=\alpha_+R^\gamma\cos(\sigma\ln R)+\alpha_-R^\gamma\sin(\sigma\ln R).
\end{equation}
A particularly simple $f(R)$ theory leading to the occurrence of a Big Rip at a final time in the future is that corresponding to $\sigma=0$. This corresponds to\footnote{There is another $f(R)$ solution 
with $\sigma=0$ and a different value of $\gamma$, that will lead to an effective behaviour for the equation of state parameter corresponding to $w=-(13+4\sqrt{6})/3$. Being this value for the effective equation of state so far away from the current observational bounds, we will simply disregard it in what follows.} 
$w=(-13+4\sqrt{6})/3\simeq-1.067$, which is a reasonable value as compared with current observational constraints for the equation of state parameter of dark energy \cite{Ade:2015xua}. 
In this case the theory is described by the following function
\begin{equation}\label{fR}
f(R)=\alpha_+\,R^\gamma,\qquad{\rm with}\qquad \gamma=2+\sqrt{3/2},
\end{equation}
where $\alpha_+$ is a constant parameter.
We will take this theory as a proxy theory when investigating the quantum realm close to the BR singularity $f(R)$-cosmology.

\section{$f(R)$ quantum cosmology}\label{sec:QC}
Alternative theories of gravity have become today a kind of paradigm
motivated by the limitations of GR to describe the cosmological evolution without the consideration of new ingredients. The understanding of these theories as fundamental classical theories demand, therefore, the
consideration of the corresponding quantum framework. On the other hand, it can be noted that $f(R)$ theories, which are formulated in the Jordan frame, have a formulation in the Einstein frame.
Although the corresponding cosmological models seem to be equivalent at the classical level when one considers a transformation of the units and restrict his/her attention to observable 
quantities \cite{Faraoni:1999hp,Capozziello:2010sc,Domenech:2015qoa,Capozziello:2011et},
the equivalence at the quantum level is much more subtle \cite{Kamenshchik:2014waa}. Hence, we consider necessary to focus on the Jordan frame formulation of the theory to design the quantisation scheme
before entering into a debate about a possible equivalence of those frameworks at the quantum level.
Therefore, in this section we first derive a point-like Lagrangian for $f(R)$ theories in the Jordan frame suitable for quantisation, in subsection \ref{sec:mL}. Then we summarise the scheme outlined by Vilenkin
in reference \cite{Vilenkin:1985md} to obtain the WDW equation for this point-like Lagrangian, in subsection \ref{sec:WdW}.

\subsection{Minisuperspace Lagrangian}\label{sec:mL}
In GR the second derivatives of the scale factor appearing in $\mathcal{L}(a,\,\dot a,\,\ddot a)$ can be removed by integration by parts. However, this is not necessarily the case
for an action of the form (\ref{L1}). As it was discussed in detail by Vilenkin in reference \cite{Vilenkin:1985md}, the standard approach to canonical quantisation in this case
consists in introducing another variable such that it allows us to remove the dependence on $\ddot a$. This will imply that the fourth-order differential equations of motion will 
be expressed as two sets of second-order differential equations.
One can choose this new variable to be the scalar curvature $R$ to express the action as $\mathcal{L}(a,\,\dot a,\,R,\,\dot R)$. However, as $R$ is not independent of $a$,
its definition has to be implemented as a constraint in equation (\ref{L1})
\begin{eqnarray}\label{L2}
 S&=&\mathcal{V}_{(3)}\int dt \, a^3\left\{f(R)\right. \nonumber\\
 &-&\left.\upsilon \left[R-6\left(\frac{\ddot a}{a}+\frac{\dot a^2}{a^2}+\frac{k}{a^2}\right)\right] \right\},
\end{eqnarray}
where we prefer not to specify yet the value of $k$. The Lagrange multiplier, $\upsilon$ can be obtained by varying the action with respect to $R$. This is
\begin{equation}\label{lambda}
 \upsilon=f_R(R).
\end{equation}
Substituting the value given by equation (\ref{lambda}) in the action and integrating by parts
the term containing $\ddot a$, one obtains the following Lagrangian
\begin{eqnarray}\label{L3}
 \mathcal{L}(a,\dot a,R,\dot R)&=&\mathcal{V}_{(3)} \left\{a^3\left[f(R)-Rf_R(R)\right] \right. \nonumber\\
 &-& \left. 6a^2f_{RR}(R)\dot a\dot R+6af_R(R)(k-\dot a^2)\right\}.
\end{eqnarray}

This point-like Lagrangian has been used in classical scenarios when studying the Noether symmetry approach to cosmology, see for example 
references \cite{Capozziello:1996bi,Capozziello:2008ima,Capozziello:2011et}.
Nevertheless, in order to consider the quantum framework, it can be useful to diagonalise the derivative part of the Lagrangian. For this purpose,
we use a change of variables qualitatively similar to that applied by Vilenkin in reference\footnote{There is a factor $\sqrt{12}$ of difference in our definition of $q$ because we prefer to use only the constant $R_0$.} 
\cite{Vilenkin:1985md}. This is
\begin{eqnarray}\label{qx}
 q=\sqrt{R_0}\,a\left(f_R/f_{R0}\right)^{1/2}\,{\rm and}\,\,
 x={\rm ln}\left(f_R/f_{R0}\right)^{1/2},
\end{eqnarray}
where we again assume the dependence of $f_R$ on $R$ and $f_{R0}\equiv f_R(R_0)$.
Note that we need a constant $R_0$ in order to consider the logarithm of a quantity with dimensions.
In reference~\cite{Vilenkin:1985md} $R_0$ is taken to be the curvature of the self-consistent de Sitter solution, given by $R_0f_{R0}-2f_0=0$, with $f_0=f(R_0)$.
Nevertheless, this choice is not always convenient as we will comment in more detail in section \ref{sec:QfR}. For the time being, let us just consider that $R_0$ is such that the transformation is well-defined
in the range of interest for the particular $f(R)$ theory.
Considering the change of variables (\ref{qx}) in the Lagrangian (\ref{L3}), one can obtain 
\begin{eqnarray}\label{L4}
 \mathcal{L}(x,\dot x,q,\dot q)&=&\mathcal{V}_{(3)} \left(\frac{R_0f_R}{f_{R0}}\right)^{-3/2}q^3  \left\{f-Rf_R\right.\nonumber\\
 &-&\left.6f_{R} \frac{\dot q^2}{q^2}
+6f_R \dot x^2+6k\frac{R_0}{f_{R0}}\frac{f_R^2}{q^2}\right\},
\end{eqnarray}
where we are now assuming $R=R(x)$ obtained from equation (\ref{qx}).

\subsection{Modified Wheeler--DeWitt equation}\label{sec:WdW}
As it was presented in reference~\cite{Vilenkin:1985md}, the WDW equation corresponding to an $f(R)$ theory for a FLRW universe 
can be obtained by quantising the Hamiltonian corresponding to Lagrangian (\ref{L4}).
Noting that 
\begin{eqnarray}
P_q&=&\frac{\partial\mathcal{L}}{\partial \dot q}=-12\,\mathcal{V}_{(3)}R_0^{-3/2}f_{R0}^{3/2}f_R^{-1/2}q\,\dot q,\\
P_x&=&\frac{\partial\mathcal{L}}{\partial \dot x}=12\,\mathcal{V}_{(3)}R_0^{-3/2}f_{R0}^{3/2}f_R^{-1/2}q^3\,\dot x,
\end{eqnarray}
the Hamiltonian can be expressed as
\begin{widetext}
\begin{equation}
 \mathcal{H}=-\mathcal{V}_{(3)}q^3\left(\frac{R_0f_R}{f_{R0}}\right)^{-3/2} \left\{ f-Rf_R+6k\frac{R_0}{f_{R0}}\frac{f_R^2}{q^2}+
 \frac{6R_0^3}{(12)^2\mathcal{V}_{(3)}^2f_{R0}^3}\frac{f_R^2}{q^4}\left[P_q^2-\frac{P_x^2}{q^2}\right]\right\}.
\end{equation}
\end{widetext}
Now, assuming the usual quantisation recipe, that is $P_q\rightarrow-i\partial_q$ and $P_x\rightarrow-i\partial_x$, we obtain a WdW equation that is equivalent to \cite{Vilenkin:1985md}
\begin{equation}\label{WdW1}
\left[\partial^2_q-\frac{1}{q^2}\partial^2_x-V(q,x)\right]\Psi(q,x)=0,
\end{equation}
where the potential is given by
\begin{equation}\label{V1}
V(q,x)=\frac{q^2}{\lambda^2}\left[k+\frac{f_{R0}}{6R_0}\left(f-Rf_R\right)\frac{q^2}{f_R^2}\right],
\end{equation}
with $\lambda=R_0/(12\mathcal{V}_{(3)}f_{R0})$.
Note that there is a factor 12 of difference between our potential (\ref{V1}) and that presented in reference \cite{Vilenkin:1985md}, that comes from the same factor of difference in the definition of $q$ in equation (\ref{qx}).
More importantly, we want to emphasize the well-known ambiguity of the theory regarding operator ordering. That is, we could have chosen a different factor ordering when obtaining equation (\ref{WdW1}), which could ultimately give rise to different wave functions. According to reference \cite{Vilenkin:1985md}, nonzero values of the factor ordering parameters introduce only unimportant modifications in the pre-exponential factor of the semi-classical wave function. 
Moreover, in this paper we want to investigate the potential avoidance of singularities by applying the DW criterion, which is independent of the factor ordering at least in some particular models \cite{Albarran:2015cda,Albarran:2017swy}. However, one should keep in mind that we are adopting an assumption that may affect the generality of our results. In particular, we are considering a ``natural" factor ordering discussed in previous literature \cite{DeWitt,Hawking1986} as providing a reasonable Hamiltonian constraint. 
On the other hand, but related with the operator ordering, when considering the Hamiltonian constraint, $H\Psi=0$, one should analyze the Hermiticity of such operator. One can find a discussion about Hermiticity of quantum contraints, for example, in section 6.3 of reference \cite{Barvinsky1} (see also section 4 of reference \cite{DeWitt}).

Before proceeding further a few words on the modified WDW equation (\ref{WdW1}) are in order as a mean to compare with what happens in GR. First of all, we notice  that the signature of the minisuperspace DeWitt metric 
\begin{equation}
G^{AB}=
\begin{bmatrix}
    1 & 0  \\
    0 & -\frac{1}{q^2} & 
\end{bmatrix}
\end{equation}
is similar to the one in GR with a matter content corresponding to a standard scalar field. Therefore, what we have proven is that this is the case even when dealing with a BR singularity. This is in strike difference to what happens in GR in presence of a BR singularity induced by a phantom minimally coupled scalar field where $G^{AB}$
has a positive signature \cite{Dabrowski:2006dd,Kamenshchik:2007zj,BouhmadiLopez:2009pu,Albarran:2015cda}.

\section{Quantum treatment of the Big Rip in $f(R)$-gravity}\label{sec:QfR}

Now, let us focus on a theory of the form given by equation (\ref{fR}) and in a cosmological model with\footnote{Please notice that the curvature term is anyway negligible as compared with the potential term when the universe approaches the BR.} $k=0$. In fact, we will assume an $f(R)$ as defined in Eq.~(\ref{fR}) with $\gamma=2+\sqrt{3/2}$.
As we want to investigate the behaviour of the model close to the BR singularity, we are mostly interested in the regime of large values of $R$.
The transformation given by equation (\ref{qx}) is well-defined in this range taking a constant $R_0$ with a small but non-vanishing value.
Nevertheless, note that the $R_0$ suggested by Vilenkin in reference \cite{Vilenkin:1985md}, that is the solution
of $R_0f_{R0}-2f_0=0$, corresponds in this case to  $R_0=0$. 
Therefore, we choose a different definition for $R_0$. In particular we take it to be of the order of the current scalar curvature of our Universe, that is $R_0\sim 4 \cdot10^4\,{\rm km}^2{\rm s}^{-2}{\rm Mpc}^{-2}$ assuming Planck data \cite{Ade:2015xua},  although we could have taken any non-vanishing value.
Note that the theory we chose, i.e. Eq.~(\ref{fR}) with $\gamma=2+\sqrt{3/2}$, is able 
to describe the evolution of our Universe in the future, when the matter content is diluted. Thus, our choice of $R_0$ implies that
$R_0<R$ for the model and, therefore, the transformation (\ref{qx}) is well-defined in the range of interest.
So, we can consider the WDW equation given by equation (\ref{WdW1})
to be suitable for our purposes, with $q$ and $x$ given by
\begin{eqnarray}\label{qx2}
 q=\sqrt{R_0}\,a\left(R/R_0\right)^{\frac{\gamma-1}{2}}\,,\,  \,\,\,\,\, x={\rm ln} \left(R/R_0\right)^{\frac{\gamma-1}{2}},
\end{eqnarray}
with $\gamma=2+\sqrt{3/2}$. Substituting the function (\ref{fR}) in the potential (\ref{V1}), we obtain
\begin{equation}\label{V2}
V(q,x)=-\frac{A}{\lambda^2}e^{-Bx}q^4,
\end{equation}
with
\begin{eqnarray}
A=\frac{\gamma-1}{6\gamma}=\frac{1}{30}(1+\sqrt{6}),
\end{eqnarray}
and
\begin{eqnarray}
B=2\frac{\gamma-2}{\gamma-1}=6-2\sqrt{6}.
\end{eqnarray}

Taking potential (\ref{V2}) into account in equation (\ref{WdW1}), the WdW equation can be written as
\begin{equation}\label{WdW2}
\left[q^2\partial^2_q-\partial^2_x+\frac{A}{\lambda^2}e^{-Bx}q^6\right]\Psi(q,x)=0.
\end{equation}
This equation could be handled in an easier way, if the potential had a dependence only in one of the variables.
In order to rewrite equation (\ref{WdW2}) in that form, we consider a change of variables of the form used in reference  \cite{Albarran:2016ewi}; i.e., 
\begin{equation}
 q=r(z)\,\varphi \qquad{\rm and}\qquad x=z.
\end{equation}
It can be seen that the potential will only depend on $\varphi$ if we take $r(z)=e^{Bz/6}$. In this case, equation (\ref{WdW2}) takes the form
\begin{widetext}
\begin{equation}\label{WdW3}
 \left[\left(1-\frac{B^2}{36}\right)\varphi^2\partial_\varphi^2-\frac{B^2}{36}\varphi\,\partial_\varphi+\frac{B}{3}\varphi\,\partial_\varphi\,\partial_z-\partial_z^2
 +\frac{A}{\lambda^2}\varphi^6\right]\Psi(\varphi,z)=0,
\end{equation}
\end{widetext}
with
\begin{equation}
 \varphi=\sqrt{R_0}\,a\left(R/R_0\right)^{\frac{2\gamma-1}{2}}\,\, {\rm and} \,\,\, z={\rm ln} \left(R/R_0\right)^{\frac{\gamma-1}{2}}.
\end{equation}

Now, we will assume that the BR singularity may be avoided if the wave function $\Psi(\varphi,z)$ satisfies the DW criterion, that is if 
$\Psi(\varphi,z)\rightarrow0$ when we approach the BR. As we do not expect the wave function to be peaked along the classical trajectory in this regime, 
$R$ and $a$ will take completely independent values on this regime. Therefore, in order to consider a region close to the BR, we should assume either $a\rightarrow\infty$
or $R\rightarrow\infty$. Both choices imply $\varphi\rightarrow\infty$, but in one case $z$ is arbitrary whereas in the other one $z\rightarrow\infty$.
We will consider that the DW criterion is satisfied if $\Psi(\varphi,z)\rightarrow0$ for $\varphi\rightarrow\infty$ and $z$ arbitrary, as it is the most general choice.

In order to solve equation (\ref{WdW3}), we will assume the following approximation:
\begin{equation}
 \varphi\,\partial_\varphi\partial_z\Psi(\varphi,z)\ll\varphi^2\partial_\varphi^2\Psi(\varphi,z)
 \end{equation}
 and
 \begin{equation}
  \varphi\,\partial_\varphi\Psi(\varphi,z)\ll\varphi^2\partial_\varphi^2\Psi(\varphi,z),
\end{equation}
when $\varphi\rightarrow\infty$. That this approximation is indeed satisfied is checked in the appendix~\ref{A1}.
Under this approximation, equation (\ref{WdW3}) can be written as
\begin{equation}\label{WdW4}
 \left[\left(1-\frac{B^2}{36}\right)\varphi^2\partial_\varphi^2-\partial_z^2
 +\frac{A}{\lambda^2}\varphi^6\right]\Psi(\varphi,z)=0.
\end{equation}
This equation can be solved assuming an ansatz for the wave function of the form \cite{Albarran:2016ewi} 
\begin{equation}
 \Psi(\varphi,z)=\sum_k a_k U_k(\varphi)C_k(z),
\end{equation}
where $a_k$ is the amplitude of each solution and $k$ is related with the energy associate with the solution (not to be confused with the spatial curvature that has been now fixed to be $0$).
Therefore, equation (\ref{WdW4}) is equivalent to the following two equations:
\begin{equation}\label{eC}
 \partial_z^2C_k(z)=k^2\,C_k(z),
\end{equation}
and
\begin{equation}\label{eU}
 \left[\left(1-\frac{B^2}{36}\right)\varphi^2\partial_\varphi^2+\frac{A}{\lambda^2}\varphi^6+k^2\right]U_k(\varphi)=0.
\end{equation}

Equation (\ref{eC}) can be easily solved. The solution for $k^2>0$ is
\begin{equation}\label{c1}
 C_k(z)= a_+e^{kz}+a_-e^{-kz},
\end{equation}
whereas for $k^2<0$ we have 
\begin{equation}\label{c2}
 C_k(z)= b_+e^{i\vert{k}\vert z}+b_-e^{-i\vert{k}\vert z},
\end{equation}
with $a_\pm$ and $b_\pm$ are constant. The solution (\ref{c1}) will be finite for any value of $z$ if $a_+=0$, whereas the solution (\ref{c2}) is always finite.

On the other hand, equation (\ref{eU}) can be approximated by
\begin{equation}\label{eU2}
 \left[\left(1-\frac{B^2}{36}\right)\varphi^2\partial_\varphi^2+\frac{A}{\lambda^2}\varphi^6\right]U_k(\varphi)=0,
\end{equation}
in the region of large values for $\varphi$. This equation can be solved by means  of Bessel functions as (cf.~Eq.~9.1.51 of Ref.~\cite{abramowitz1})
\begin{equation}
U_k(\varphi)=
\varphi^{1/2}\left[U_1\,J_{\frac16}\left(\frac13\tilde\lambda\varphi^3
\right)+U_2\,Y_{\frac16}\left(\frac13\tilde\lambda\varphi^3
\right)\right],
\end{equation}
where
\begin{equation}
\tilde\lambda^2=\frac{A}{\left(1-\frac{B^2}{36}\right)} \frac{1}{\lambda^{2}},
\end{equation}
$U_1$ and $U_2$ are arbitrary constants, and $J_\nu(z)$ and $Y_\nu(z)$ are the Bessel functions of the first and second kind, respectively.
Given that we are very close to the regime of large values of $\varphi$, the solution can be further approximated as (cf.~Eq.~9.2.1 of Ref.~\cite{abramowitz1})
\begin{eqnarray}\label{sol}
 U_k(\varphi)&=&\sqrt{\frac{6}{\tilde\lambda\pi}}\frac{1}{\varphi}\left[U_1\cos\left(\frac{\tilde\lambda}{3}\varphi^3-\frac{\pi}{3}\right)\right.\nonumber\\
 &+&\left.U_2\sin\left(\frac{\tilde\lambda}{3}\varphi^3-\frac{\pi}{3}\right)\right].
\end{eqnarray}
Therefore, we can conclude that $U_k(\varphi)\rightarrow0$ when $\varphi\rightarrow\infty$. Taking into account that $C_k(z)$ remains finite for any value of $z$ with our choice $a_+=0$,
$\Psi(\varphi,z)\rightarrow0$  when $\varphi\rightarrow\infty$. 
There are also solutions of the wave function that do not vanish at the BR. 
However, as imposing that the wave function vanishes at the singular boundary is not inconsistent, we can follow DeWitt spirit and argue that those; i.e. with a non-vanishing wave function at the singularity, are not physical solutions, as the probability to reach that boundary should be zero\footnote{If investigations on other aspects of this model could point towards the need of taking the dismissed solutions into account, then one would conclude that the DeWitt criterion is inconsistent in this case.}.
Consequently, it turns out that the DW criterion is satisfied pointing towards singularity avoidance.

\section{Discussion}\label{sec:conc.}

The observational data currently available show that the accelerated expansion of our Universe is compatible with the existence of a phantom fluid, which could point towards the occurrence of a future BR singularity. That cosmic evolution can also be modeled by alternative theories of gravity without the introduction of exotic cosmic components, in particular this can be done in the framework of $f(R)$ theories of gravity. As it is well known, the classical theory of gravity, be it GR or $f(R)$-gravity, is not able to predict what happens at (or even close to) that singularity. Therefore, in this paper we have considered the formulation of $f(R)$ quantum cosmology (by considering an $f(R)$ formulation as a fundamental theory) to study the potential avoidance of the BR.

We have quantized an specific $f(R)$ cosmological model predicting a BR singularity in the framework of quantum geometrodynamics. That is, following the procedure outlined in reference~\cite{Vilenkin:1985md}, we have developed a canonical quantisation restricting our analysis to the minisuperspace to formulate the modified WDW equation.
When quantizing we have explicitly taken a particular factor ordering and implicitly assumed an Hermitian Hamiltonian.
An interesting technical point that deserves to be emphasized is that the signature of the minisuperspace DeWitt metric is similar to that obtained in GR when assuming an ordinary (non-phantom) scalar field independently of the particular $f(R)$ theory under investigation. Therefore, the WDW equation is hyperbolic even if the quantised universe has a BR as its classical fate.

We have focused our attention on a particularly simple $f(R)$ theory of gravity leading to a BR singularity and compatible with current observational constraints. Then, we have considered the modified WDW equation of that particular theory and, after developing some approximations justified in the appendix, assumed an ansatz commonly used in the literature for the form of the wavefunction of the universe. However, it should be emphasized that in our case this ansatz is not splitting the matter and the geometric part of the wavefunction, as both minisuperspace variables are related to geometry in our case. Then, we have obtained the solutions to the WDW equation and showed that the DW criterion for singularity avoidance can be imposed, without any further analysis of the family of solutions to the equation. It should be noted, however, that that the DW criterion is based on the existence of a consistent probability interpretation of the wavefunction that is still unknown. Hence, our results hint towards the avoidance of the BR in metric $f(R)$ theories, although a definitive answer will require a formulation of that probability interpretation. 

Finally, we want to emphasize that in order to obtain a complete formulation (and interpretation) of the quantum system, one needs to be able to construct a Hilbert space of solutions for the WDW equation  equipped with an inner product. We have not discussed the potential definition of that Hilbert space and its inner product in the model investigated in the present paper. It should be noted that such inner product is necessary for discussing the physical significance of the solutions of the WDW equation. In particular, it is the common root for our assumptions regarding the symmetric character of the Hamiltonian constraint and the applicability of the DeWitt criterion. This construction is not possible in general but it is for some particular cosmological models \cite{Barvinsky1, Kamens1, Barvinsky2, Kamens2,ana1}. We postpone, however, the corresponding detailed analysis for the particular model considered in this paper for a future research.

\section*{Acknowledgments}
Dedicated to the memory of Prof.~Pedro F.~Gonz\'alez-D\'iaz (our former PhD supervisor). 
A.~A.-S.~is funded by the Alexander von Humboldt Foundation. A.~A.-S.~work is also supported by the Project. No.~\mbox{MINECO FIS2017-86497-C2-2-P} from Spain, and partially by the grant \mbox{GACR-14-37086G} of the Czech Science Foundation.
The research of M.~B.-L.~is supported by the Basque Foundation of Science Ikerbasque. She also would like to acknowledge the partial support from the Basque government Grant No. IT956-16 (Spain) and FONDOS FEDER under grant FIS2017-85076-P (MINECO/AEI/FEDER, UE). She also wishes to thank the hospitality of the Universidad Complutense of Madrid where part of this work was carried out.
The work of P.~M.-M.~has been supported by the projects FIS2014-52837-P (Spanish MINECO) and FIS2016-78859-P (AEI/FEDER, UE).
This article is based upon work from COST Action CA15117 ``Cosmology and Astrophysics Network for Theoretical Advances and Training Actions (CANTATA)'', supported by COST (European Cooperation in Science and Technology)


\appendix

\section{Validity of the approximation}\label{A1}

We next proof the validity of the approximation we used to solve the modified WdW equation.

Taking into account equation (\ref{sol}), we can see that the approximations taken for equation (\ref{WdW3}) implies that
\begin{equation}
 \varphi^2\partial_\varphi^2\Psi\sim \varphi^5\gg\varphi\partial_\varphi\Psi\sim\varphi^2,
\end{equation}
and
\begin{equation}
 \varphi^2\partial_\varphi^2\Psi\sim \varphi^5\gg\varphi\partial_\varphi\partial_z\Psi\sim\varphi^2.
\end{equation}
On the other hand, those inequality have to be satisfied at large $\varphi$ where the BR singularity is reached. As can be seen easily those inequalities are indeed fulfilled on that regime.



\begin{thebibliography}{99}


\bibitem{Ade:2015xua}
  P.~A.~R.~Ade {\it et al.} [Planck Collaboration],
  ``Planck 2015 results. XIII. Cosmological parameters'',
  Astron.\ Astrophys.\  {\bf 594} (2016) A13
[\href{https://arxiv.org/abs/astro-ph/0004075}{astro-ph/0004075}].
 
\bibitem{Sahni:1999gb}
 V.~Sahni and A.~A.~Starobinsky,
 ``The Case for a positive cosmological Lambda term'',
 Int.\ J.\ Mod.\ Phys.\ D {\bf 9} (2000) 373
 [\href{https://arxiv.org/abs/astro-ph/9904398}{astro-ph/9904398}].
 
\bibitem{Weinberg:1988cp}
 S.~Weinberg,
 ``The Cosmological Constant Problem'',
 Rev.\ Mod.\ Phys.\ {\bf 61} (1989) 1.
 
\bibitem{Carroll:2000fy}
 S.~M.~Carroll,
 ``The Cosmological constant'',
 Living Rev.\ Rel.\ {\bf 4} (2001) 1
 [\href{https://arxiv.org/abs/astro-ph/0004075}{astro-ph/0004075}].

\bibitem{Padmanabhan:2002ji}
 T.~Padmanabhan,
 ``Cosmological constant: The Weight of the vacuum'',
 Phys.\ Rept.\ {\bf 380} (2003) 235
 [\href{https://arxiv.org/abs/hep-th/0212290}{hep-th/0212290}].

\bibitem{Caldwell:1999ew}
  R.~R.~Caldwell,
  ``A Phantom menace?'',
  Phys.\ Lett.\ B {\bf 545} (2002) 23
[\href{https://arxiv.org/abs/astro-ph/9908168}{astro-ph/9908168}].

\bibitem{Starobinsky:1999yw}
  A.~A.~Starobinsky,
  ``Future and origin of our universe: Modern view'',
  Grav.\ Cosmol.\  {\bf 6} (2000) 157
[\href{https://arxiv.org/abs/astro-ph/9912054}{astro-ph/9912054}].

\bibitem{Caldwell:2003vq}
  R.~R.~Caldwell, M.~Kamionkowski and N.~N.~Weinberg,
  ``Phantom energy and cosmic doomsday'',
  Phys.\ Rev.\ Lett.\  {\bf 91} (2003) 071301
[\href{https://arxiv.org/abs/astro-ph/0302506}{astro-ph/0302506}].  
\bibitem{BouhmadiLopez:2004me}
  M.~Bouhmadi-L\'opez and J.~A.~Jim\'enez Madrid,
 ``Escaping the big rip?'',
  JCAP {\bf 0505} (2005) 005
  [\href{https://arxiv.org/abs/astro-ph/0404540}{astro-ph/0404540}].
\bibitem{BouhmadiLopez:2006fu}
  M.~Bouhmadi-L\'opez, P.~F.~Gonz\'alez-D\'iaz and P.~Mart\'in-Moruno,
  ``Worse than a big rip?'',
  Phys.\ Lett.\ B {\bf 659} (2008) 1
  [\href{https://arxiv.org/abs/gr-qc/0612135}{gr-qc/0612135}].

\bibitem{BouhmadiLopez:2007qb}
  M.~Bouhmadi-L\'opez, P.~F.~Gonz\'alez-D\'iaz and P.~Mart\'in-Moruno,
  ``On the generalised Chaplygin gas: Worse than a big rip or quieter than a sudden singularity?'',
  Int.\ J.\ Mod.\ Phys.\ D {\bf 17} (2008) 2269
 [\href{https://arxiv.org/abs/0707.2390}{arXiv:0707.2390 [gr-qc]}].
\bibitem{MCP}
M.~Bouhmadi-L\'opez, C. Kiefer and P.~Mart\'in-Moruno,  
work in progress. 


\bibitem{Wheeler}
J.~A.~Wheeler, 
``On the nature of quantum geometrodynamics'',
 Ann. Phys. {\bf 2} (1957) 604.

\bibitem{DeWitt}
B.~S.~DeWitt,  
``Quantum theory of gravity. I. The canonical theory'',
Phys. Rev {\bf 160} (1967) 1113.

\bibitem{Halliweell}
J.~J.~Halliwell, 
``Introductory lectures on quantum cosmology'', in { \it Quantum cosmology and baby universes}
edited by S.~Coleman, J.~B.~Hartle, T.~Piran and S.~Weinberg (World Scientic, London, UK, 1990).

\bibitem{Misner}
C.~Misner, ``Minisuperspace'', in {\it Magic without magic: John Archibald Wheeler}, edited by J.~R.~Klauder (Freeman, San Francisco, USA, 1972).

\bibitem{Kuchar}
K.~V.~Kuchar and M.~P.~Ryan, ``Is minisuperspace quantization valid?: Taub in mixmaster'',
Phys. Rev. D {\bf 40} (1989) 3982.


\bibitem{clausbook} C. Kiefer, {\textit Quantum Gravity}. Third edition (Oxford University Press, Oxford, 2012).



\bibitem{Capozziello:2011et}
  S.~Capozziello and M.~De Laurentis,
  ``Extended Theories of Gravity'',
  Phys.\ Rept.\  {\bf 509} (2011) 167
[\href{https://arxiv.org/abs/1108.6266}{arXiv:1108.6266 [gr-qc]}]. 
  
\bibitem{Sotiriou:2008rp}
  T.~P.~Sotiriou and V.~Faraoni,
  ``f(R) Theories Of Gravity'',
  Rev.\ Mod.\ Phys.\  {\bf 82} (2010) 451
[\href{https://arxiv.org/abs/0805.1726}{arXiv:0805.1726 [gr-qc]}]. 

\bibitem{Nojiri:2010wj}
  S.~Nojiri and S.~D.~Odintsov,
  ``Unified cosmic history in modified gravity: from F(R) theory to Lorentz non-invariant models'',
  Phys.\ Rept.\  {\bf 505} (2011) 59
 [\href{https://arxiv.org/abs/1011.0544}{arXiv:1011.0544 [gr-qc]}]. 


\bibitem{Bouhmadi-Lopez:2016dcf}
  M.~Bouhmadi-L\'opez and C.~Y.~Chen,
 ``Towards the Quantization of Eddington-inspired-Born-Infeld Theory'',
  JCAP {\bf 1611} (2016) no.11,  023
[\href{https://arxiv.org/abs/1609.00700}{arXiv:1609.00700 [gr-qc]}].  

\bibitem{Albarran:2017swy}
  I.~Albarran, M.~Bouhmadi-L\'opez, C.~Y.~Chen and P.~Chen,
  ``Doomsdays in a modified theory of gravity: A classical and a quantum approach'',
  Phys.\ Lett.\ B {\bf 772} (2017) 814
[\href{https://arxiv.org/abs/1703.09263}{arXiv:1703.09263 [gr-qc]}]. 

\bibitem{Nojiri:2008fk}
  S.~Nojiri and S.~D.~Odintsov,
  ``The Future evolution and finite-time singularities in F(R)-gravity unifying the inflation and cosmic acceleration'',
  Phys.\ Rev.\ D {\bf 78} (2008) 046006
  doi:10.1103/PhysRevD.78.046006
  [arXiv:0804.3519 [hep-th]].


\bibitem{Bamba:2008ut}
  K.~Bamba, S.~Nojiri and S.~D.~Odintsov,
  ``The Universe future in modified gravity theories: Approaching the finite-time future singularity'',
  JCAP {\bf 0810} (2008) 045
  [arXiv:0807.2575 [hep-th]].
  
\bibitem{Capozziello:2009hc}
  S.~Capozziello, M.~De Laurentis, S.~Nojiri and S.~D.~Odintsov,
  ``Classifying and avoiding singularities in the alternative gravity dark energy models'',
  Phys.\ Rev.\ D {\bf 79} (2009) 124007
  [arXiv:0903.2753 [hep-th]].

\bibitem{Nojiri:2017ncd}
  S.~Nojiri, S.~D.~Odintsov and V.~K.~Oikonomou,
  ``Modified Gravity Theories on a Nutshell: Inflation, Bounce and Late-time Evolution'',
  Phys.\ Rept.\  {\bf 692} (2017) 1
  [\href{https://arxiv.org/abs/1705.11098}{arXiv:1705.11098 [gr-qc]}]. 
  
\bibitem{Morais:2015ooa}
J.~Morais, M.~Bouhmadi-L\'opez and S.~Capozziello,
``Can $f(R)$ gravity contribute to (dark) radiation?'',
JCAP {\bf 1509} (2015) no.09,  041.
[\href{https://arxiv.org/abs/1507.02623}{arXiv:1507.02623 [gr-qc]}]. 

\bibitem{Faraoni:1999hp}
  V.~Faraoni and E.~Gunzig,
  ``Einstein frame or Jordan frame?'',
  Int.\ J.\ Theor.\ Phys.\  {\bf 38} (1999) 217
 [\href{https://arxiv.org/abs/astro-ph/9910176}{astro-ph/9910176}].

\bibitem{Capozziello:2010sc}
  S.~Capozziello, P.~Mart\'in-Moruno and C.~Rubano,
  ``Physical non-equivalence of the Jordan and Einstein frames'',
  Phys.\ Lett.\ B {\bf 689} (2010) 117
[\href{https://arxiv.org/abs/1003.5394}{arXiv:1003.5394[gr-qc]}]. 

\bibitem{Domenech:2015qoa}
  G.~Dom\`enech and M.~Sasaki,
  ``Conformal Frame Dependence of Inflation'',
  JCAP {\bf 1504} (2015) no.04,  022
[\href{https://arxiv.org/abs/1501.07699}{arXiv:1501.07699[gr-qc]}]. 



\bibitem{Kamenshchik:2014waa}
  A.~Y.~Kamenshchik and C.~F.~Steinwachs,
  ``Question of quantum equivalence between Jordan frame and Einstein frame'',
  Phys.\ Rev.\ D {\bf 91} (2015) no.8,  084033
[\href{https://arxiv.org/abs/1408.5769}{arXiv:1408.5769[gr-qc]}]. 


\bibitem{Vilenkin:1985md}
A.~Vilenkin,
``Classical and Quantum Cosmology of the Starobinsky Inflationary Model'',
Phys.\ Rev.\ D {\bf 32} (1985) 2511.

\bibitem{Capozziello:1996bi}
  S.~Capozziello, R.~De Ritis, C.~Rubano and P.~Scudellaro,
  ``Noether symmetries in cosmology'';
  Riv.\ Nuovo Cim.\  {\bf 19N4} (1996) 1.

\bibitem{Capozziello:2008ima}
  P.~Mart\'in-Moruno, S.~Capozziello and C.~Rubano,
  ``Dark energy and dust matter phases from an exact $f(R)$-cosmology model'',
  Phys.\ Lett.\ B {\bf 664} (2008) 12
[\href{https://arxiv.org/abs/0804.4340}{arXiv:0804.4340[gr-qc]}].

\bibitem{Albarran:2015cda}
  I.~Albarran, M.~Bouhmadi-L\'opez, F.~Cabral and P.~Mart\'in-Moruno,
``The quantum realm of the "Little Sibling" of the Big Rip singularity'',
  JCAP {\bf 1511} (2015) no.11,  044
[\href{https://arxiv.org/abs/1509.07398}{arXiv:1509.07398[gr-qc]}]. 



\bibitem{Hawking1986}
S. W. Hawking and D. N. Page, 
Nucl.\ Phys.\ B {\bf 264} (1986) 185.


\bibitem{Dabrowski:2006dd}
  M.~P.~D\c{a}browski, C.~Kiefer and B.~Sandh\"ofer,
  ``Quantum phantom cosmology'',
  Phys.\ Rev.\ D {\bf 74} (2006) 044022
[\href{https://arxiv.org/abs/hep-th/0605229}{hep-th/0605229}].  

  
\bibitem{Kamenshchik:2007zj}
  A.~Kamenshchik, C.~Kiefer and B.~Sandh\"ofer,
  ``Quantum cosmology with big-brake singularity'',
  Phys.\ Rev.\ D {\bf 76} (2007) 064032
 [\href{https://arxiv.org/abs/0705.1688}{arXiv:0705.1688[gr-qc]}].

  
\bibitem{BouhmadiLopez:2009pu}
  M.~Bouhmadi-L\'opez, C.~Kiefer, B.~Sandh\"ofer and P.~Vargas Moniz,
  ``On the quantum fate of singularities in a dark-energy dominated universe'',
  Phys.\ Rev.\ D {\bf 79} (2009) 124035
[\href{https://arxiv.org/abs/0905.2421}{arXiv:0905.2421[gr-qc]}].


\bibitem{Albarran:2016ewi}
I.~Albarran, M.~Bouhmadi-L\'opez, C.~Kiefer, J.~Marto and P.~Vargas Moniz,
``Classical and quantum cosmology of the little rip abrupt event'',
Phys.\ Rev.\ D {\bf 94} (2016) no.6,  063536.
[\href{https://arxiv.org/abs/1604.08365}{arXiv:1604.08365[gr-qc]}].

\bibitem{abramowitz1}
M. Abramowitz and I. Stegun,
\textit{Handbook of Mathematical Functions} (Dover, 1980).


 \bibitem{Barvinsky1}
 A.~O.~Barvinsky,
  ``Unitarity approach to quantum cosmology'',
  Phys.\ Rept.\  {\bf 230} (1993) 237.

 \bibitem{Kamens1}
 A.~Y.~Kamenshchik and S.~Manti, 
 ``Classical and quantum Big Brake cosmology for scalar field and tachyonic models'',
 Phys. Rev. D {\bf 85} (2012) 123518.
[\href{https://arxiv.org/abs/1202.0174v2}{arXiv:1202.0174v2[gr-qc]}]
 
  \bibitem{Kamens2}
 A.~Y.~Kamenshchik,
 ``Quantum cosmology and late-time singularities'', 
 Class. Quant. Grav. { \bf 30} (2013) 173001.
 [\href{https://arxiv.org/abs/1307.5623v2}{arXiv:1307.5623v2 [gr-qc]}]
 
  \bibitem{Barvinsky2}
 A. O. Barvinsky and A. Y. Kamenshchik, 
 ``Selection rules for the Wheeler-DeWitt equation in quantum cosmology'',
 Phys. Rev. D {\bf 89} (2014) 4, 043526.
 [\href{https://arxiv.org/abs/1312.3147v2}{arXiv:1312.3147v2 [gr-qc]}]

\bibitem{ana1}
A.~Alonso-Serrano,  L.~J.~Garay, G.~A.~Mena Marug\'an, 
``Correlations across horizons in quantum cosmology'',
Phys.\ Rev.\ D {\bf 90} (2014) 12,  124074. 
[\href{https://arxiv.org/abs/1410.2543v2}{arXiv:1410.2543v2[gr-qc]}]



\end{thebibliography}
\end{document}